\documentclass[prd,nopacs,preprintnumbers,twocolumn,amsmath,nofootinbib,amssymb]{revtex4}
\usepackage{graphicx,color,dcolumn,booktabs,bm,makecell}
\usepackage{longtable,lscape}
\usepackage{txfonts}
\usepackage{overpic}
\usepackage{epstopdf}
\usepackage{indentfirst}
\usepackage{feynmf}   
\usepackage{slashed}  
\usepackage{cases}
\usepackage{color}
\usepackage{soul}
\usepackage{cancel}
\usepackage{float}
\usepackage{makecell}
\usepackage{gensymb}
\usepackage{multirow}
\usepackage{epsfig,dsfont,amssymb,amsmath,amsfonts,amsbsy,mathrsfs}

\graphicspath{{Figures/}} %

\usepackage{hyperref}
\hypersetup{colorlinks,citecolor=blue,anchorcolor=red,menucolor=red,linkcolor=red,filecolor=red,runcolor=red,urlcolor=blue,frenchlinks=red}

\makeatletter
\@addtoreset{equation}{section}
\makeatother
\renewcommand{\theequation}{\arabic{section}.\arabic{equation}}
\allowdisplaybreaks

\newcolumntype{I}{!{\vrule width 0.9pt}}

\begin{document}

\title{Higher strangeonium decays into light flavor baryon pairs like $\Lambda\bar{\Lambda}$, $\Sigma\bar{\Sigma}$, and $\Xi\bar{\Xi}$}
\author{Zi-Yue Bai$^{1,2,3,5}$}\email{baizy15@lzu.edu.cn}
\author{Qin-Song Zhou$^{2,4,5,6}$}\email{zhouqs13@lzu.edu.cn}
\author{Xiang Liu$^{1,2,3,4,5}$\footnote{Corresponding author}}\email{xiangliu@lzu.edu.cn}

\affiliation{$^1$School of Physical Science and Technology, Lanzhou University, Lanzhou 730000, China\\
$^2$Lanzhou Center for Theoretical Physics, Key Laboratory of Theoretical Physics of Gansu Province, Lanzhou University, Lanzhou 730000, China\\
$^3$Key Laboratory of Quantum Theory and Applications of MoE, Lanzhou University,
Lanzhou 730000, China\\
$^4$MoE Frontiers Science Center for Rare Isotopes, Lanzhou University, Lanzhou 730000, China\\
$^5$Research Center for Hadron and CSR Physics, Lanzhou University and Institute of Modern Physics of CAS, Lanzhou 730000, China\\
$^6$School of Physical Science and Technology, Inner Mongolia University, Hohhot 010021, China}

\begin{abstract}
In this work, we investigate the decay behaviors of several higher strangeonia into $\Lambda\bar\Lambda$ through a hadronic loop mechanism, enabling us to predict some physical observables, including the branching ratios. Furthermore, we assess the reliability of our research by successfully reproducing experimental data related to the cross section of $e^+e^-\to\Lambda\bar{\Lambda}$ interactions. In this context, we account for the contributions arising from higher strangeonia, specifically $\phi(4S)$ and $\phi(3D)$. Additionally, we extend this study to encompass higher strangeonia decays into other light flavor baryon pairs, such as $\Sigma\bar\Sigma$ and $\Xi\bar\Xi$. By employing the same mechanism, we aim to gain valuable insights into the decay processes involving these particles. By conducting this investigation, we hope to shed light on the intricate decay mechanisms of higher strangeonia and their interactions with various baryons pairs.
\end{abstract}

\pacs{} %
\maketitle

\section{introduction}\label{sec1}

Studying hadron spectroscopy has emerged as an effective approach to enhance our understanding of non-perturbative quantum chromodynamics (QCD), particularly with the recent observation of a series of new hadronic states. These include the $XYZ$ charmoniumlike states and the $P_c/P_{cs}$ pentaquark states, which have generated considerable interest in the research community (for recent progress, refer to review articles \cite{Liu:2013waa,Esposito:2016noz,Chen:2016spr,Guo:2017jvc,Liu:2019zoy,Brambilla:2019esw,Chen:2022asf,Chen:2016qju}).
Among these intriguing hadronic states, the strangeoniumlike $Y(2175)$ \cite{BaBar:2006gsq,Belle:2008kuo,BES:2007sqy,BESIII:2014ybv} has garnered significant attention. The discovery of the $Y(2175)$ state has not only spurred investigations into exotic states such as hybrid states \cite{Ding:2006ya} and tetraquarks \cite{Agaev:2019coa,Xin:2022qnv,Wang:2006ri,Agaev:2020zad,Drenska:2008gr,Deng:2010zzd}, but it has also provided an opportunity to explore higher strangeonium states \cite{,Barnes:2002mu,Pang:2019ttv,Wang:2012wa,Ding:2007pc}. Distinguishing between the various assignments for the $Y(2175)$ has now become a crucial task. Consequently, studying its decay behaviors, as one facet of hadron spectroscopic behavior, can yield valuable insights into its internal structure.

Recently, the BESIII Collaboration conducted a measurement of the cross section for the process $e^+e^-\to \Lambda\bar{\Lambda}$ at center-of-mass energy from the production threshold up to 3.0 GeV \cite{BESIII:2023ioy}. This energy range coincides with the masses of the $Y(2175)$ state and several higher strangeonia that have been predicted \cite{Barnes:2002mu,Pang:2019ttv,Wang:2021gle,Wang:2021abg}. Naturally, this intriguing discovery has piqued our curiosity, prompting further investigation into the decays of higher strangeonia states into light flavor baryon pairs such as $\Lambda\bar{\Lambda}$.

Unlike the typical OZI-allowed two-body strong decays with mesonic final states, the decay of the higher strangeonium into $\Lambda\bar{\Lambda}$ proceeds through the creation of two pairs of quarks and antiquark from the vacuum, as illustrated by the quark pair creation model \cite{Micu:1968mk,LeYaouanc:1977gm,Xiao:2019qhl}. This distinctive decay mechanism sets it apart from other decay processes.

For higher strangeonia, the unquenched effect cannot be ignored. To quantitatively depict the decays of higher strangeonia into light flavor baryon pairs, we propose the utilization of hadronic loops composed of kaons and nucleons as a bridge connecting the initial higher strangeonium states to the final light flavor baryon pairs. Introducing the hadronic loop mechanism provides a realistic representation of the unquenched effect.
In this study, we calculate the branching ratio of higher strangeonia decays into $\Lambda\bar{\Lambda}$ and compare it with experimental data obtained from the process of $e^+e^-\to \Lambda\bar{\Lambda}$. Additionally, our investigation reveals the potential contribution of higher strangeonia to the cross section of $e^+e^-\to \Lambda\bar{\Lambda}$.
Furthermore, we examine the decays involving other light flavor baryon pairs, such as $\Sigma\bar\Sigma$ and $\Xi\bar\Xi$, and predict their respective branching ratios. These predictions can be tested in future experiments, offering opportunities for further exploration in this field.

This paper is organized as follows. After the Introduction, in Sec. \ref{sec2}, we illustrate the detailed calculation of higher strangeonium decays into $\Lambda\bar{\Lambda}$, which can be quantitatively depicted by the hadronic loop mechanism. And then, we present the numerical results of these discussed decays, and do a simple fit to the cross section date of $e^+e^-\to\Lambda\bar\Lambda$, which can be applied to show the reliability of the discussed decay mechanism in this work.
In Sec. \ref{sec3}, we further predict the branching ratios of some higher strangeonia decays into  $\Sigma\bar\Sigma$ and $\Xi\bar\Xi$. Finally, the paper ends with a summary.

\section{Investigation of the  strangeonium decays into  $\Lambda\bar{\Lambda}$ via the hadronic loop mechanism}\label{sec2}

\subsection{Hadronic loop mechanism}

The unquenched effect is important for us to understand the spectra and the decay behaviors of hadrons, especially when we face the excited hadrons approaching threshold \cite{Luo:2019qkm,Zhang:2022pxc,Duan:2020tsx,Duan:2021bna,Duan:2021alw,Liu:2015taa,Zhao:2020qpd}. The hadronic loop mechanism is a quantitative description of the unquenched effect, which has been successfully employed to explain the anomalous hidden-charm/bottom decay behaviors of higher charmonia/bottomonia \cite{Cheng:2004ru,Liu:2006dq,Liu:2009dr,Meng:2007tk,Meng:2008dd,Meng:2008bq,Chen:2011qx,Chen:2011zv,Chen:2011pv,Chen:2011jp,Chen:2014ccr,Wang:2015xsa,Wang:2016qmz,Huang:2017kkg,Zhang:2018eeo,Huang:2018cco,Huang:2018pmk,Li:2021jjt,Bai:2022cfz,Li:2022leg}, where the predicted branching ratios are usually close to the experimental measurements. It also has been applied to study the decay behaviors of higher charmonia into a pair of charmed or strange baryons \cite{Qian:2021gby,Qian:2021neg}. In this section, we utilize the hadronic loop mechanism to study the partial decay widths of higher strangeonia to $\Lambda\bar{\Lambda}$, where 
hadronic loops can be as a bridge 
connecting the initial higher strangeonia to final state $\Lambda\bar{\Lambda}$.

As shown in Refs. \cite{Wang:2021gle,Wang:2021abg}, these higher strangeonia have abundant decay modes, with some channels like $KK_1(1270)$, $KK^*(1410)$ also dominating in the higher strangeonium decays. Under the hadronic loop mechanism, we should consider their contributions to the higher strangeonium decays into a light flavor baryon pair. In fact, for completeness, all hadronic loops that can connect the initial and final states should be included. However, in a realistic calculation we have to face the  serious problem that we do not know the couplings of the some involved kaons like $K^*(1410)$ with nucleon and $\Lambda/\Sigma/\Xi$. It makes this strategy become uncontrolled. To give a  quantitative calculation, we have to adopt another strategy, where the hadroinc loops are restricted to a subset, as applied in the following study. By fitting some experimental data, we can obtain the corresponding range of parameters like $\alpha$, by which we can further give other theoretical predictions. We call this treatment ``{\it a desperate attempt, but an effective way}". In our earlier work, we have adopted this strategy to deal with some concrete issues. For example, we predicted the $\Upsilon(10860)\to\Upsilon(1^3D_J\eta)$ \cite{Wang:2016qmz} process by hadronic loop mechanism, which was later confirmed by the Belle Collaboration \cite{Belle:2018hjt}.

\begin{figure}[htbp]\centering
  \includegraphics[width=86mm]{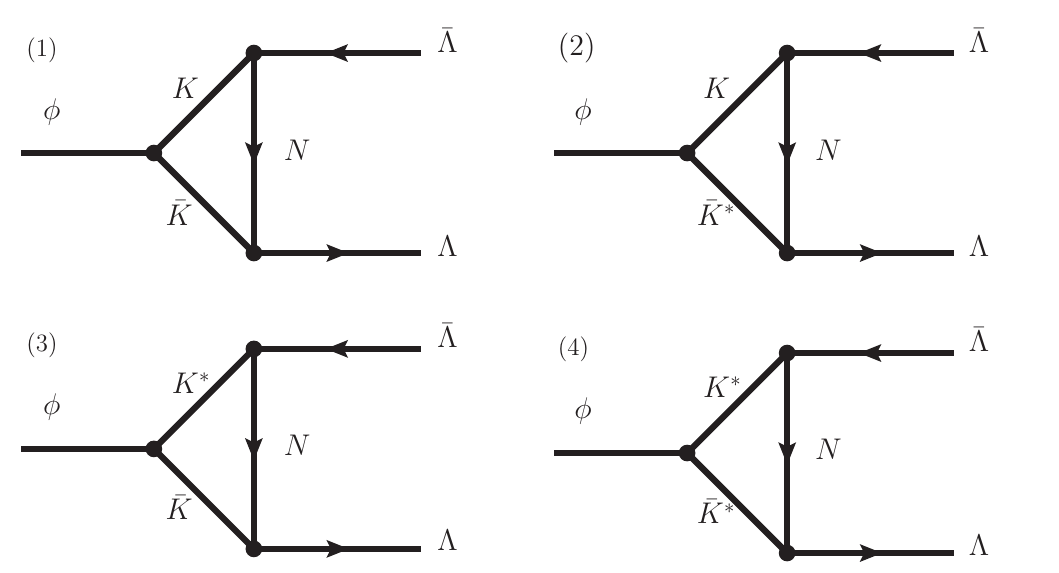}\\
  \caption{The schematic diagrams illustrating the higher strangeonium decays into $\Lambda\bar{\Lambda}$ via the hadronic loop mechanism.}
\label{fig:Lambda}
\end{figure}

{In this work, we consider two subsets of hadronic loops. The first one is that we consider $1S$ states ($K$ and $K^*$) of the kaon to construct the hadronic loops, as shown in Fig. \ref{fig:Lambda}.}
Concretely, in our calculation, we employ the effective Lagrangian approach to describe the interaction vertices shown in Fig. \ref{fig:Lambda}. This approach allows us to effectively capture the dynamics of the involved interactions, and the concrete calculations rely on the following effective Lagrangians \cite{Chen:2011cj,Wang:2021gle,Rijken:2006en,Nagels:2015lfa}:
\begin{equation}
\begin{split}
\mathcal{L}_{\phi KK}&=ig_{\phi KK}(\bar K\partial_\mu K-\partial_\mu\bar KK)\phi^\mu,\\
\mathcal{L}_{\phi KK^*}&=g_{\phi KK^*}\varepsilon_{\mu\nu\rho\sigma}(\bar K\partial^\rho K^{*\sigma}+\partial^\rho\bar K^{*\sigma} K)\partial^\mu\phi^\nu,\\
\mathcal{L}_{\phi K^*K^*}&=ig_{\phi K^*K^*}(({\bar K}_\nu^*\overset\leftrightarrow\partial_\mu K^{*\nu})\phi^\mu+\bar K^{*\mu}(K^*_\nu\overset\leftrightarrow\partial_\mu\phi^\nu)\\
&+(\phi^\nu\overset\leftrightarrow\partial_\mu\bar K^*_\nu)K^{*\mu}),\\
\mathcal{L}_{\mathcal{B}_1\mathcal{B}_2K}&=\frac{if_{\mathcal{B}_1\mathcal{B}_2K}}{m_{\pi^+}}\bar{\mathcal{B}}_1\gamma_\mu\gamma_5\mathcal{B}_2\partial^\mu K,  \\
\mathcal{L}_{\mathcal{B}_1\mathcal{B}_2K^*}&=g_{\mathcal{B}_1\mathcal{B}_2K^*}\bar{\mathcal{B}}_1\gamma_\mu\mathcal{B}_2K^{*\mu}\\
&+\frac{if_{\mathcal{B}_1\mathcal{B}_2K^*}}{4m_p}\bar{\mathcal{B}}_1\sigma_{\mu\nu}\mathcal{B}_2(\partial^\mu K^{*\nu}-\partial^\nu K^{*\mu}),
\end{split}
\end{equation}
with $\sigma_{\mu\nu}=\frac{i}{2}[\gamma_\mu,\gamma_\nu]$. Here, the $\phi$, $K^{(*)}$, and  $\mathcal{B}_1/\mathcal{B}_2$ denote the fields of the corresponding mesons and baryons. $N$ represent the exchanged proton or neutron. The masses of the charged pion and proton are denoted as $m_{\pi^+}=139.6\,\text{MeV}$ and $m_p=938.3\,\text{MeV}$, respectively. With these preparations, we can now write down the decay amplitudes for the higher strangeonium decays into $\Lambda\bar\Lambda$, i.e.,
\begin{equation}
\begin{split}
\mathcal{M}_{(1)}^{\Lambda\bar{\Lambda}}=&i^3\int\frac{dq^4}{(2\pi)^4}g_{\phi KK}\epsilon_\phi^\mu(q_{1\mu}-q_{2\mu})\left(\frac{f_{\Lambda NK}}{m_{\pi^+}}\right)^2  \\
&\times\bar u(p_2)\slashed{q}_2\gamma^5(\slashed q+m_N)\slashed q_1\gamma^5v(p_1)\\
&\times\frac{1}{q_1^2-m_K^2}\frac{1}{q_2^2-m_K^2}\frac{1}{q^2-m_N^2}\mathcal{F}^{2}(q^2,m_N^2),
\end{split}
\end{equation}

\begin{equation}
\begin{split}
\mathcal{M}_{(2)}^{\Lambda\bar{\Lambda}}=&-i^3\int\frac{dq^4}{(2\pi)^4}g_{\phi KK^*}\varepsilon_{\mu\nu\alpha\beta}\epsilon_\phi^\mu p^\nu q_2^\alpha\bar u(p_2)\Big(g_{\Lambda NK^*}\gamma_\rho \\
&+\frac{f_{\Lambda NK^*}}{2m_p}q_2^\lambda\sigma_{\lambda\rho}\Big)(\slashed q+m_N)\frac{f_{\Lambda NK}}{m_{\pi^+}}\slashed q_1\gamma^5v(p_1)\\
&\times\frac{1}{q_1^2-m_K^2}\frac{-g^{\beta\rho}+q_2^\beta q_2^\rho\slash{m_{K^{*}}^2}}{q_2^2-m_{K^{*}}^{2}}\frac{1}{q^2-m_N^2}\mathcal{F}^{2}(q^2,m_N^2),\\
\end{split}
\end{equation}

\begin{equation}
\begin{split}
\mathcal{M}_{(3)}^{\Lambda\bar{\Lambda}}=&-i^3\int\frac{dq^4}{(2\pi)^4}g_{\phi KK^*}\varepsilon_{\mu\nu\alpha\beta}\epsilon_\phi^\mu p^\nu q_1^\alpha\bar u(p_2)\frac{f_{\Lambda NK}}{m_{\pi^+}}\slashed q_2\gamma^5\\
&\times(\slashed q+m_N)\Big(g_{\Lambda NK^*}\gamma_\rho+\frac{f_{\Lambda NK^*}}{2m_p}q_1^\lambda\sigma_{\lambda\rho}\Big)v(p_1)\\
&\times\frac{-g^{\beta\rho}+q_1^\beta q_1^\rho\slash{m_{K^{*}}^2}}{q_1^2-m_{K^{*}}^{2}}\frac{1}{q_2^2-m_K^{2}}\frac{1}{q^2-m_N^2}\mathcal{F}^{2}(q^2,m_N^2),\\
\end{split}
\end{equation}

\begin{equation}
\begin{split}
\mathcal{M}_{(4)}^{\Lambda\bar{\Lambda}}=&i^3\int\frac{dq^4}{(2\pi)^4}g_{\phi K^*K^*}\epsilon_\phi^\mu\Big(g_{\alpha\beta}(q_{1\mu}-q_{2\mu})+g_{\beta\mu}(p_\alpha+q_{2\alpha})\\
&-g_{\alpha\mu}(p_\beta+q_{1\beta})\Big)\bar u(p_2)\Big(g_{\Lambda NK^*}\gamma_\rho+\frac{f_{\Lambda NK^*}}{2m_p}q_2^\lambda\sigma_{\lambda\rho}\Big)\\
&\times(\slashed q+m_N)\Big(g_{\Lambda NK^*}\gamma_\eta+\frac{f_{\Lambda NK^*}}{2m_p}q_1^\kappa\sigma_{\kappa\eta}\Big)v(p_1) \\
&\times\frac{-g^{\alpha\eta}+q_1^\alpha q_1^\eta\slash{m_{K^{*}}^2}}{q_1^2-m_{K^{*}}^{2}}\frac{-g^{\beta\rho}+q_2^\beta q_2^\rho\slash{m_{K^{*}}^2}}{q_2^2-m_{K^{*}}^{2}}\frac{1}{q^2-m_N^2}\\&\times\mathcal{F}^{2}(q^2,m_N^2),
\end{split}
\end{equation}
where the dipole form factor $\mathcal{F}(q^2,m_E^2)=\left(\frac{m_E^2-\Lambda^2}{q^2-\Lambda^2}\right)^2$ is utilized to describe the off shell effect of the exchanged baryon and ensure the convergence of the loop integrals \cite{Qian:2021gby,Qian:2021neg}. In this context, $m_E$ and $q$ represent the mass and four-momentum of the exchanged baryon, respectively. The cutoff parameter $\Lambda$ can be parametrized as $\Lambda = m_E + \alpha \Lambda_{\text{QCD}}$, where $\Lambda_{\text{QCD}} = 220$ MeV, and the value of $\alpha$ is chosen to be of order 1 to ensure that $\Lambda$ is in proximity to the mass of the exchanged baryon $m_E$ \cite{Cheng:2004ru}.

The total decay amplitude for the $\phi\to\Lambda\bar{\Lambda}$ process can be expressed as
\begin{equation}
\mathcal{M}_{\text{total}}^{\Lambda\bar{\Lambda}} = 2\left(\mathcal{M}_{(1)}^{\Lambda\bar{\Lambda}} + \mathcal{M}_{(2)}^{\Lambda\bar{\Lambda}} + \mathcal{M}_{(3)}^{\Lambda\bar{\Lambda}} + \mathcal{M}_{(4)}^{\Lambda\bar{\Lambda}}\right),
\end{equation}
where the factor of 2 in the total decay amplitude arises from the summation over the isospin doublets  $(K^{(*)},K^{(*)+})$ and $(n,p)$.

Then, the decay width of $\phi\to\Lambda\bar{\Lambda}$ process can be calculated by
\begin{equation}
\begin{split}
\Gamma=&\frac{1}{3}\frac{|\vec p_1|}{8\pi m_\phi^2}\sum_{{\rm spin}}\Big|\mathcal{M}_{{\rm total}}^{\Lambda\bar{\Lambda}}\Big|^2,\\
\end{split}
\end{equation}
where the factor ${1}/{3}$ and the summation $\sum_{\rm spin}$ are from the average of the polarizations of the initial $\phi$ state and the sum of all possible spins of the final states $\Lambda$ and $\bar{\Lambda}$, respectively.

We first present the numerical results of the branching ratios for the decays of the $\phi(4S)$ and $\phi(3D)$ into $\Lambda\bar{\Lambda}$, considering the $\alpha$ dependence. Subsequently, we perform a fitting analysis of the cross section for $e^+e^-\to\Lambda\bar{\Lambda}$, incorporating the contributions from the $\phi(4S)$ and $\phi(3D)$.
By this way, the reliability of the adopted hadronic loop mechanism can be further tested.  
Finally, using the same approach, we will predict the decay behaviors of possible strangeonia located above the $\Lambda\bar{\Lambda}$ threshold. Specifically, we will study their decays into $\Sigma\bar{\Sigma}$ and $\Xi\bar{\Xi}$, while constraining the range of $\alpha$.

\subsection{Branching ratios of  the $\phi(4S)$ and $\phi(3D)$ decays into $\Lambda\bar\Lambda$}

In this subsection, we present the calculation results of the $\phi(4S)$ and $\phi(3D)$ decays into $\Lambda\bar{\Lambda}$ final states, focusing specifically on the strangeonia that lie above the $\Lambda\bar{\Lambda}$ threshold. To facilitate our study, we  utilize the resonance parameters of these strangeonium states, which have been obtained from Refs. \cite{Wang:2021gle,Wang:2021abg}.

\begin{table}[htbp]
\centering
\caption{The resonance parameters and di-electron widths of the involved strangeonia above the $\Lambda\bar\Lambda$ threshold predicted by Refs. \cite{Wang:2021gle,Wang:2021abg}.}
\renewcommand\arraystretch{1.3}
\begin{tabular*}{86mm}{l@{\extracolsep{\fill}}ccc}
\toprule[1pt]
\toprule[0.5pt]
States               &Mass ($\text{GeV}$)      &Width ($\text{MeV}$)     &$\Gamma_{\phi}^{e^+e^-}\, (\text{eV})$    \\
\midrule[0.5pt]
$\phi(4S)$          &2.423                   &140                     &0.049                      \\
$\phi(3D)$          &2.519                   &171                     &0.010                      \\
$\phi(5S)$          &2.671                   &104                     &0.029                      \\
$\phi(4D)$          &2.744                   &129                     &0.006                      \\
$\phi(6S)$          &2.871                   &71                      &0.016                      \\
$\phi(5D)$          &2.924                   &106                     &0.004                      \\
\bottomrule[0.5pt]
\bottomrule[1pt]
\label{table:resonance}
\end{tabular*}
\end{table}


Before showing the results, we illustrate the procedure used to determine the coupling constants in the decay amplitudes. To achieve this, we start by obtaining the coupling constants associated with the $\phi K^{(*)}K^{(*)}$ vertices. These coupling constants can be fixed by ensuring agreement with the partial decay widths of the $\phi\to K^{(*)}K^{(*)}$ process, as reported in Refs. \cite{Wang:2021gle,Wang:2021abg}.
In Table \ref{coupling Constants}, we collect the partial decay widths obtained from the aforementioned theoretical works, along with the corresponding coupling constants that we have fixed based on this information.

\begin{table}[htbp]
\centering
\caption{The partial decay widths of $\phi\to K^{(*)}K^{(*)}$ and the corresponding fixed coupling constants.}
\label{coupling Constants}
\renewcommand\arraystretch{1.3}
\begin{tabular*}{86mm}{l@{\extracolsep{\fill}}cc}
\toprule[1pt]
\toprule[0.5pt]
Channal    &Partial widths (MeV)    &Coupling constants \\
\midrule[0.5pt]
$\phi(4S)\to KK$     &6.58         &0.520 \\
$\phi(4S)\to KK^*$   &20.30        &0.451 $\text{GeV}^{-1}$ \\
$\phi(4S)\to K^*K^*$ &1.96         &0.061 \\
$\phi(5S)\to KK$     &4.68         &0.407 \\
$\phi(5S)\to KK^*$   &11.13        &0.270 $\text{GeV}^{-1}$\\
$\phi(5S)\to K^*K^*$ &0            &0 \\
$\phi(6S)\to KK$     &3.10         &0.312\\
$\phi(6S)\to KK^*$   &6.67         &0.181 $\text{GeV}^{-1}$\\
$\phi(6S)\to K^*K^*$ &0            &0 \\
$\phi(3D)\to KK$     &14.36        &0.745\\
$\phi(3D)\to KK^*$   &8.55         &0.269 $\text{GeV}^{-1}$\\
$\phi(3D)\to K^*K^*$ &23.77        &0.185\\
$\phi(4D)\to KK$     &9.29         &0.562\\
$\phi(4D)\to KK^*$   &4.64         &0.165 $\text{GeV}^{-1}$\\
$\phi(4D)\to K^*K^*$ &16.64        &0.120\\
$\phi(5D)\to KK$     &6.68         &0.455\\
$\phi(5D)\to KK^*$   &2.86         &0.114 $\text{GeV}^{-1}$\\
$\phi(5D)\to K^*K^*$ &14.73        &0.082\\
\bottomrule[0.5pt]
\bottomrule[1pt]
\end{tabular*}
\end{table}

The coupling constants associated with the $\Lambda NK^{(*)}$ vertices are adopted from the results obtained by the extended-soft-core (ESC) model \cite{Nagels:2015lfa}, as displayed below: 
$f_{\Lambda NK}=-0.950$, $g_{\Lambda NK^*}=-3.557$, and
$f_{\Lambda NK^*}=-14.935$.

\begin{figure}[htbp]\centering
  \includegraphics[width=86mm]{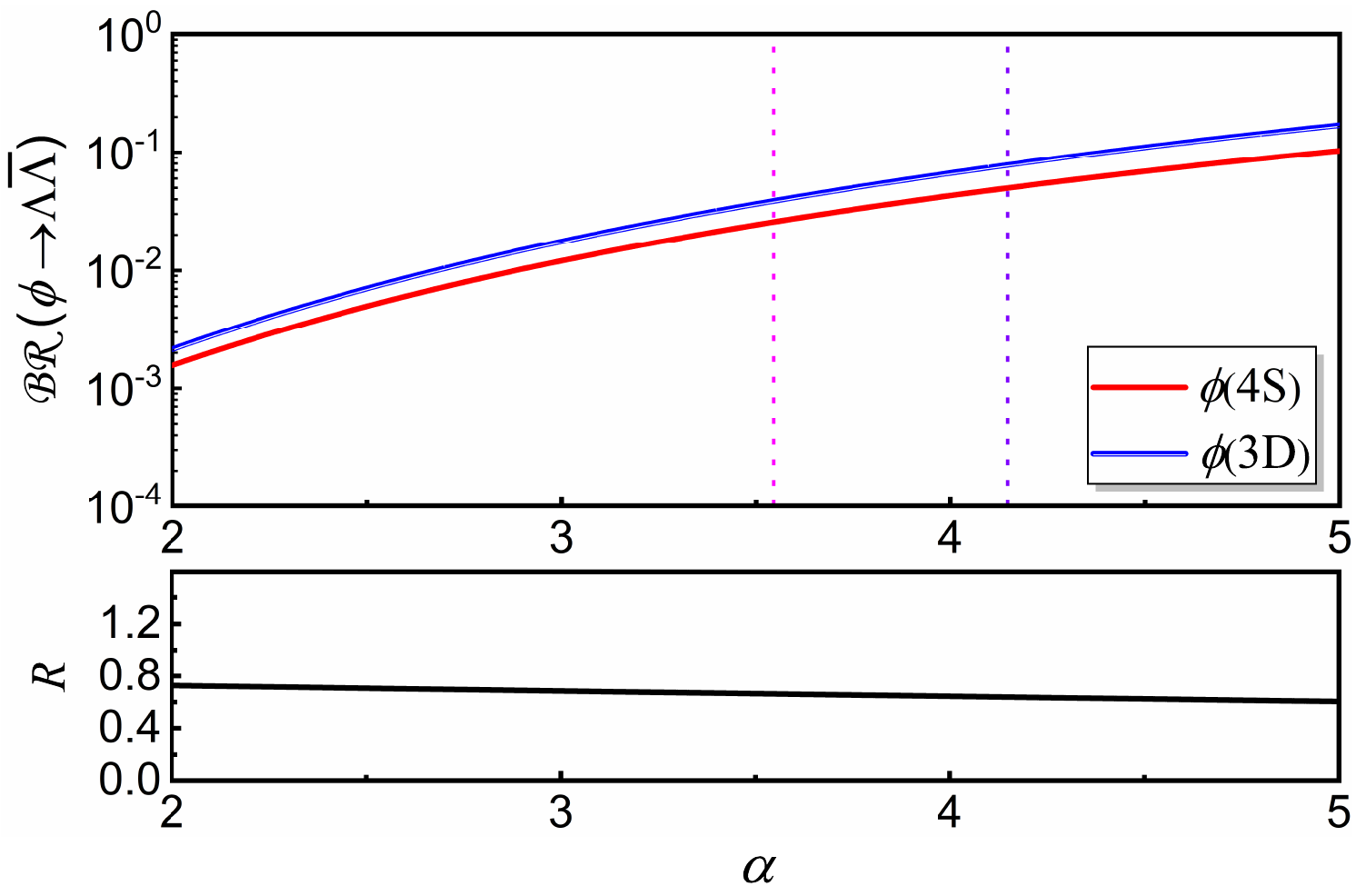}\\
  \caption{(Color online.) The $\alpha$ dependence of the branching ratios for the $\phi(4S)\to\Lambda\bar\Lambda$ (red solid curve) and $\phi(3D)\to\Lambda\bar\Lambda$ (blue solid thick-thin curve) decays is shown in the top plot. Additionally, the bottom plot (black solid curve) illustrates the ratio of these branching ratios. The violet dotted vertical line and the magenta dotted vertical line represent the maximum and central values of $\alpha$, respectively, which were obtained from the total fit to the data of $e^+e^-\to\Lambda\bar\Lambda$, as displayed in Fig. \ref{fig:fit}.}
\label{fig:br}
\end{figure}

The branching ratios for the decays $\phi(4S)\to\Lambda\bar{\Lambda}$ and $\phi(3D)\to\Lambda\bar{\Lambda}$, incorporating the dependence on the parameter $\alpha$ from the form factor, are presented in Fig. \ref{fig:br}. In the range of $\alpha$ spanning from 2.0 to 5.0, the ratio between the two branching ratios, $R=\mathcal{BR}(\phi(4S)\to\Lambda\bar{\Lambda})/\mathcal{BR}(\phi(3D)\to\Lambda\bar{\Lambda})$, varies between 0.605 and 0.728. Notably, this ratio exhibits a gradual decrease with increasing values of $\alpha$, eventually reaching an approximately constant value.

Moving forward, our next step involves combining these results with the experimental cross section for $e^+e^-\to\Lambda\bar{\Lambda}$. By performing a comparative analysis, we aim to constrain the parameter $\alpha$, which is applied to predict the branching ratios of higher strangeonia decays into $\Sigma\bar{\Sigma}$ and $\Xi\bar{\Xi}$.

\subsection{A simple fit to the cross section of $e^+e^-\to\Lambda\bar\Lambda$}

Recently, the BESIII Collaboration conducted a new measurement of the cross section for the $e^+e^-\to\Lambda\bar\Lambda$ process \cite{BESIII:2023ioy}. Prior reports on this process were made by the BaBar \cite{BaBar:2007fsu} and BESIII Collaborations \cite{BESIII:2017hyw, BESIII:2019nep}. Notably, the data accumulation was observed around 2.4 GeV, showing the possibility of higher strangeonia $\phi(4S)$ and $\phi(3D)$ contributions due to their masses being approximately 2.4 GeV \cite{BESIII:2023ioy,Wang:2021gle,Wang:2021abg}.

Motivated by these findings, our work aims to provide some insights by employing a simple fit to the experimental data. Specifically, we take into account the contributions from the $\phi(4S)$ and $\phi(3D)$ resonances, in addition to a nonresonance background. Through this analysis, we may shed light on the presence and impact of these resonances in the observed cross section data.

In concrete study, we adopt a phase space corrected Breit-Wigner distribution
\begin{equation}
\mathcal{A}_R^{\phi}=\frac{\sqrt{12\pi\Gamma_\phi^{e^+e^-}\mathcal{BR}(\phi\to\Lambda\bar\Lambda)\Gamma_\phi}}{s-m_\phi^2+im_\phi\Gamma_\phi}\sqrt{\frac{\Phi(s)}{\Phi(m_\phi^2)}}
\end{equation}
to describe the contribution from the $\phi(4S)$ and $\phi(3D)$ resonances, where $\Phi(s)$ is the two-body phase space of final states, $m_\phi$, $\Gamma_\phi$, $\Gamma_\phi^{e^+e^-}$ are the mass, width, and dielectron width of the involved higher strangeonia, respectively.

The nonresonance contribution is parametrized by a simple exponential form
\begin{equation}
\mathcal{A}_{NR}=a\,e^{-b(\sqrt{s}-2m_\Lambda)},
\label{non}
\end{equation}
where $a$ and $b$ are as the fitting parameters.

The total amplitude for the process $e^+e^-\to\Lambda\bar\Lambda$ can be written as
\begin{equation}
\mathcal{A}=\mathcal{A}_{NR}+\mathcal{A}_R^{\phi(4S)}e^{i\theta_{\phi(4S)}}+\mathcal{A}_R^{\phi(3D)}e^{i\theta_{\phi(3D)}},
\label{tot}
\end{equation}
and total cross section is 
\begin{equation}
\sigma(s) = |\mathcal{A}|^2,
\end{equation}
where $\theta_{\phi(4S)}$ and $\theta_{\phi(3D)}$ are the phase angles. Since the ratio $R$ in Fig. \ref{fig:br} remains stable, we adopt a typical value of $\mathcal{BR}(\phi(4S)\to\Lambda\bar\Lambda)=0.66\times\,\mathcal{BR}(\phi(3D)\to\Lambda\bar\Lambda)$ in the fitting procedure.

\begin{table}[htbp]
\centering
\caption{The fitted parameters to the cross section data of $e^+e^-\to\Lambda\bar\Lambda$.}
\label{fit parameter}
\renewcommand\arraystretch{1.3}
\begin{tabular*}{86mm}{l@{\extracolsep{\fill}}cc}
\toprule[1pt]
\toprule[0.5pt]
Parameters    &Values    &Error $(\pm)$ \\
\midrule[0.5pt]
$a$  &$6.89\times10^{-4}$ $\text{GeV}^{-1}$  &$1.33\times10^{-4}$ $\text{GeV}^{-1}$ \\
$b$  &3.14 $\text{GeV}^{-1}$  &1.10 $\text{GeV}^{-1}$ \\
$\theta_{\phi(4S)}$  &1.73 rad  &0.36 rad \\
$\theta_{\phi(3D)}$  &0   &0.19 rad \\
${\mathcal{BR}(\phi(3D)\to\Lambda\bar\Lambda)}$  &0.039  &0.039 \\
\bottomrule[0.5pt]
\bottomrule[1pt]
\end{tabular*}
\end{table}

\begin{figure}[htbp]\centering
  \includegraphics[width=85mm]{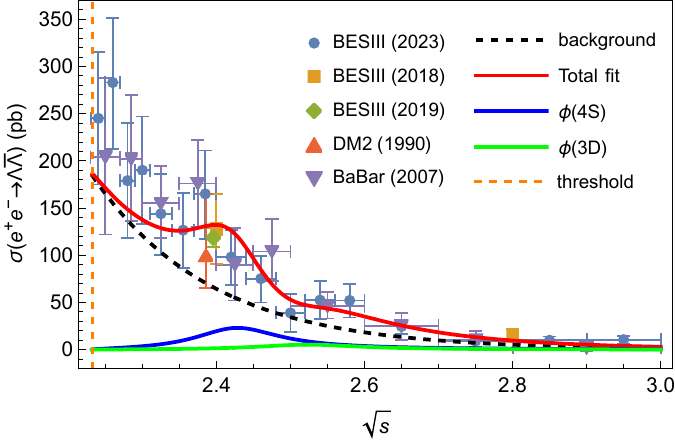}\\
  \caption{The fit to the experimental data \cite{BESIII:2023ioy,BaBar:2007fsu,BESIII:2017hyw,BESIII:2019nep,DM2:1990tut} of the cross section of $e^+e^-\to\Lambda\bar\Lambda$ supported by the spectroscopy of higher strangeonia.}
\label{fig:fit}
\end{figure}

Considering the peculiar threshold enhancement effect \cite{Haidenbauer:2016won, Cao:2018kos, Yang:2019mzq, Baldini:2007qg, Salnikov:2023zyt, Qian:2022whn}, the data from the $\Lambda\bar\Lambda$ threshold to 2.28 GeV is not included in the fit. The fit parameters are summarized in Table \ref{fit parameter}, and the fit result is presented in Fig. \ref{fig:fit}, with a obtained fit quality of $\chi^2/d.o.f=0.91$. The central and maximum values of the fitted $\mathcal{BR}(\phi(3D)\to\Lambda\bar\Lambda)$ are 0.039 and 0.078, respectively, as shown in Table \ref{fit parameter}.

Upon comparison with our results from the hadronic loop mechanism, we obtain the corresponding central and maximum values of the $\alpha$ parameter to be 3.55 and 4.15, as shown in Fig. \ref{fig:br}. Notably, the fitted branching ratios of $\phi(4S,3D)\to\Lambda\bar\Lambda$ align well with the corresponding branching ratios predicted by the hadronic loop mechanism. This observation indicates the contribution of the hadronic loop mechanism to the decays of higher strangeonia into strange baryons pairs.

Furthermore, based on these studies, we can explore the decays of other higher strangeonia into $\Lambda\bar\Lambda$, as well as other strange baryons pairs such as $\Sigma\bar\Sigma$ and $\Xi\bar\Xi$, with a reasonable $\alpha$ parameter in the range of $3.0\sim4.0$. Here, the calculated branching ratios of strangeonia above $\Lambda\bar\Lambda$ threshold decays into $\Lambda\bar\Lambda$ with $\alpha$ parameter from 3.0 to 4.0 are
\begin{equation}
\begin{split}
\mathcal{BR}(\phi(4S)\to\Lambda\bar\Lambda)=(1.21-4.31)\times10^{-2}, \\
\mathcal{BR}(\phi(3D)\to\Lambda\bar\Lambda)=(1.76-6.68)\times10^{-2}, \\
\mathcal{BR}(\phi(5S)\to\Lambda\bar\Lambda)=(6.77-23.8)\times10^{-3}, \\
\mathcal{BR}(\phi(4D)\to\Lambda\bar\Lambda)=(1.16-4.47)\times10^{-2}, \\
\mathcal{BR}(\phi(6S)\to\Lambda\bar\Lambda)=(4.73-16.8)\times10^{-3}, \\
\mathcal{BR}(\phi(5D)\to\Lambda\bar\Lambda)=(7.59-29.5)\times10^{-3}. \\
\end{split}
\label{branchratio}
\end{equation}

\subsection{The numerical result of taking a different subset for the hadronic loop: The $\phi(4S,3D)\to \Lambda\bar{\Lambda}$ case}\label{d}

\begin{figure}[htbp]\centering
  \includegraphics[width=86mm]{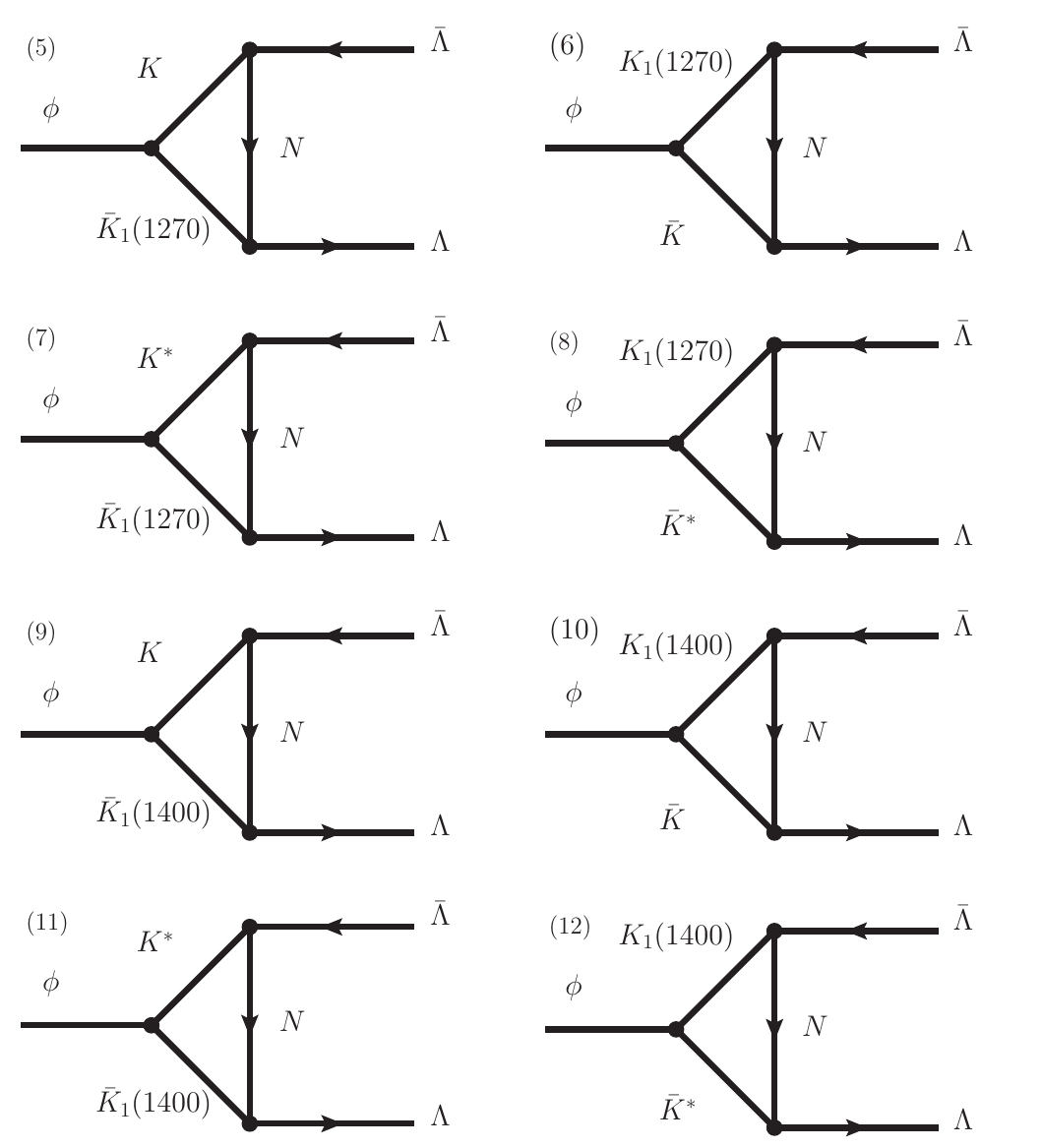}\\
  \caption{The schematic diagrams of the higher strangeonium decays into $\Lambda\bar\Lambda$ via the hadronic loops relevant to the $K_1(1270)/K_1(1400)$ states.}
\label{fig:K1}
\end{figure}

In this subsection, we adopt another subset of the hadronic loops to study the decay of $\phi(4S,3D)\to\Lambda\bar\Lambda$. The loops include not only the contribution of $1S$ states ($K$ and $K^*$) as shown in Fig. \ref{fig:Lambda}, but also $1P$ kaon states ($K_1(1270)$ and $K_1(1400)$) as depicted in Fig. \ref{fig:K1}. In order to quantitatively calculate the contributions involving the hadronic loops relevant to the $1P$ kaon states, we need information of Lagrangians and the corresponding coupling constants for the $\phi K^{(*)}K_1$ and $N\Lambda K_1$ interactions.

The effective Lagrangians of the $\phi K^{(*)}K_1$ interactions are \cite{Wang:2021gle,Kaymakcalan:1983qq,Kochelev:1999zf,Zhou:2022wwk,Liu:2022yrt} 
\begin{equation}
\begin{split}
\mathcal{L}_{\phi KK_1}&=g_{\phi KK_1}(\bar{K}K_1^\mu+\bar{K}_1^\mu K)\phi_\mu,\\
\mathcal{L}_{\phi K^*K_1}&=ig_{\phi K^*K_1}\varepsilon^{\mu\nu\alpha\beta}((\partial_\mu K^*_\alpha\partial^\lambda\partial_\lambda\phi_\nu-\partial^\lambda\partial_\lambda K^*_\alpha\partial_\mu\phi_\nu)\bar{K}_{1\beta}\\
&-(\partial_\mu \bar{K}^*_\alpha\partial^\lambda\partial_\lambda\phi_\nu-\partial^\lambda\partial_\lambda \bar{K}^*_\alpha\partial_\mu\phi_\nu)K_{1\beta}).\\
\end{split}
\end{equation}
Here, the partial decay widths and the fixed coupling constants are displayed in Table \ref{coupling Constants1}.

\begin{table}[htbp]
\centering
\caption{The partial decay widths of $\phi(4S,3D)\to K^{(*)}K_1(1270)/K^{(*)}K_1(1400)$ \cite{Wang:2021gle} and the corresponding fixed coupling constants.}
\label{coupling Constants1}
\renewcommand\arraystretch{1.3}
\begin{tabular*}{86mm}{l@{\extracolsep{\fill}}cc}
\toprule[1pt]
\toprule[0.5pt]
Channal    &Partial widths (MeV)    &Coupling constants \\
\midrule[0.5pt]
$\phi(4S)\to KK_1(1270)$     &21.56         &0.939 $\text{GeV}$\\
$\phi(4S)\to K^*K_1(1270)$ &2.52         &0.033 $\text{GeV}^{-2}$\\
$\phi(3D)\to KK_1(1270)$     &50.10         &1.413 $\text{GeV}$\\
$\phi(3D)\to K^*K_1(1270)$   &16.25        &0.070 $\text{GeV}^{-2}$\\
$\phi(4S)\to KK_1(1400)$     &3.78          &0.430 $\text{GeV}$\\
$\phi(4S)\to K^*K_1(1400)$ &17.64         &0.118 $\text{GeV}^{-2}$\\
$\phi(3D)\to KK_1(1400)$     &2.57         &0.347 $\text{GeV}$\\
$\phi(3D)\to K^*K_1(1400)$   &0.51       &0.016 $\text{GeV}^{-2}$\\
\bottomrule[0.5pt]
\bottomrule[1pt]
\end{tabular*}
\end{table}

We treat $K_1(1200)$ and $K_1(1400)$ as the mixture of the $1^1P_1$ and $1^3P_1$ kaon states \cite{Pang:2017dlw}, to work out the $N\Lambda K_1$ vertices,
\begin{equation}
\left(
\begin{array}{c}
|K_1(1270)\rangle \\
|K_1(1400)\rangle
\end{array}
\right)
=
\left(
\begin{array}{cc}
\text{cos}\,\theta & \text{sin}\,\theta \\
-\text{sin}\,\theta & \text{cos}\,\theta
\end{array}
\right)
\left(
\begin{array}{c}
|1^1P_1\rangle \\
|1^3P_1\rangle
\end{array}
\right),
\end{equation}
where $\theta=45\degree$ is taken from Ref. \cite{Pang:2017dlw}.

For the $K(1^1P_1)$ ($K_1$) and $K(1^3P_1)$ ($K_1^\prime$) exchange, the Lagrangians of the corresponding $N\Lambda K_1^{(\prime)}$ interactions have the form as following \cite{Nagels:2015lfa},
\begin{equation}
\begin{split}
&\mathcal{L}_{\mathcal{B}_1\mathcal{B}_2K_1}=\frac{f_{\mathcal{B}_1\mathcal{B}_2K_1}}{m_{b_1}}\bar{\mathcal{B}}_1\sigma_{\mu\nu}\gamma_5\mathcal{B}_2\partial_\nu K_1^\mu,\\
\mathcal{L}_{\mathcal{B}_1\mathcal{B}_2K^\prime_1}&=g_{\mathcal{B}_1\mathcal{B}_2K^\prime_1}\bar{\mathcal{B}}_1\gamma_\mu\gamma_5\mathcal{B}_2K^{\prime\mu}_1+\frac{f_{\mathcal{B}_1\mathcal{B}_2K^\prime_1}}{m_p}\bar{\mathcal{B}}_1\gamma_5\mathcal{B}_2\partial_{\mu}K^{\prime\mu}_1.
\end{split}
\end{equation}
Here, $m_{b_1}=1229.5\,\text{MeV}$ and $m_p=938.3\,\text{MeV}$ are the masses of $b_1(1235)$ and the proton, respectively. The coupling constants are written as $f_{N\Lambda K_1}$=8.325, $g_{N\Lambda K_1^\prime}$=2.954, and $f_{N\Lambda K_1^\prime}$=5.972.

\begin{figure}[htbp]\centering
  \includegraphics[width=86mm]{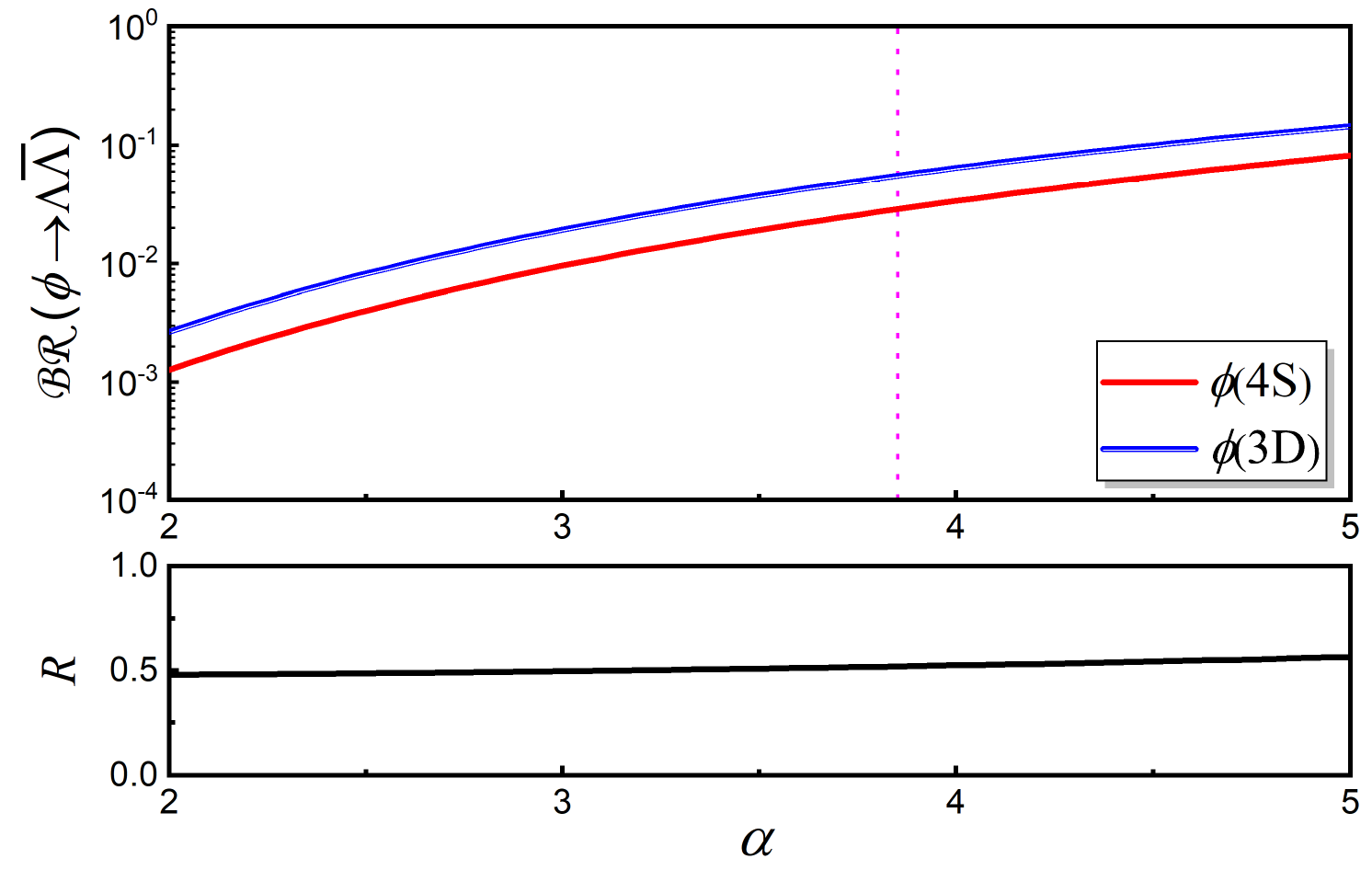}\\
  \caption{(Color online.) The $\alpha$ dependence of the branching ratios for the $\phi(4S)\to\Lambda\bar\Lambda$ (red solid curve) and $\phi(3D)\to\Lambda\bar\Lambda$ (blue solid thick-thin curve) decays by taking a new subset of the hadronic loops is shown in the top plot. In addition, the lower plot (black solid curve) shows the ratio of these branching ratios. The magenta dotted vertical line represents the central value of $\alpha$, which was obtained from the overall fit to the data of $e^+e^-\to\Lambda\bar\Lambda$, as displayed in Fig. \ref{fig:fit1}.}
\label{fig:br1}
\end{figure}

With the above preparations, now we can calculate the branching ratios of $\phi(4S)/\phi(3D)\to\Lambda\bar\Lambda$ including the contributions of hadronic loops in Fig. \ref{fig:Lambda} and Fig. \ref{fig:K1}. Concrete amplitudes are shown in Appendix. The branching ratios with the dependence of parameter $\alpha$ are listed in Fig. \ref{fig:br1}. $R=\mathcal{BR}(\phi(4S)\to\Lambda\bar{\Lambda})/\mathcal{BR}(\phi(3
D)\to\Lambda\bar{\Lambda})$, varies between 0.48 and 0.56 with $\alpha$ in the range of 2.0
$\sim$5.0.

\begin{figure}[htbp]\centering
  \includegraphics[width=85mm]{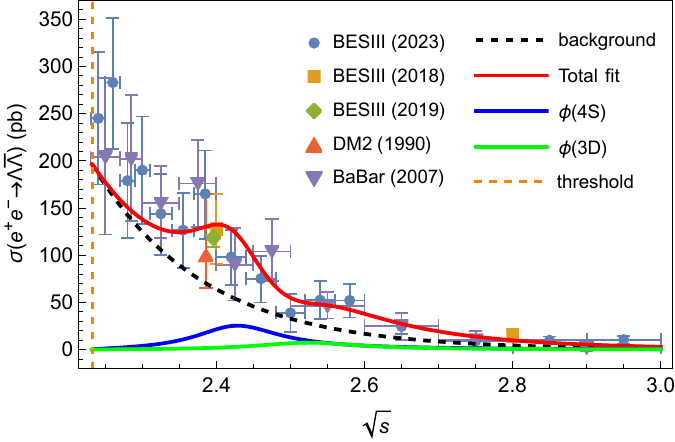}\\
  \caption{The fit to the experimental data \cite{BESIII:2023ioy,BaBar:2007fsu,BESIII:2017hyw,BESIII:2019nep,DM2:1990tut} of the cross section of $e^+e^-\to\Lambda\bar\Lambda$ by adopting a different subset for the hadronic loop.}
\label{fig:fit1}
\end{figure}

Similarly, we fit the cross section data of $e^+e^-\to\Lambda\bar\Lambda$ again with a typical value of $R=0.52$, the form of the nonresonance contribution and the total amplitude are identical to Eq. (\ref{non}) and Eq. (\ref{tot}). The fit parameters are summarized in Table \ref{fit parameter1}, and the fit result is presented in Fig. \ref{fig:fit1}, with an obtained fit quality of $\chi^2/d.o.f=0.72$.

\begin{table}[htbp]
\centering
\caption{The fitted parameters to the cross section data of $e^+e^-\to\Lambda\bar\Lambda$ with a different subset of hadronic loops.}
\label{fit parameter1}
\renewcommand\arraystretch{1.3}
\begin{tabular*}{86mm}{l@{\extracolsep{\fill}}cc}
\toprule[1pt]
\toprule[0.5pt]
Parameters    &Values    &Error $(\pm)$ \\
\midrule[0.5pt]
$a$  &$7.11\times10^{-4}$ $\text{GeV}^{-1}$  &$1.62\times10^{-4}$ $\text{GeV}^{-1}$ \\
$b$  &3.34 $\text{GeV}^{-1}$  &1.65 $\text{GeV}^{-1}$ \\
$\theta_{\phi(4S)}$  &1.63 rad  &0.43 rad \\
$\theta_{\phi(3D)}$  &0   &0.35 rad \\
${\mathcal{BR}(\phi(3D)\to\Lambda\bar\Lambda)}$  &0.054  &0.070 \\
\bottomrule[0.5pt]
\bottomrule[1pt]
\end{tabular*}
\end{table}

When restricting the parameter $\alpha$ in the range of $3.4\sim4.4$, according to the fit results shown in Table \ref{fit parameter1} and Fig. \ref{fig:br1}, the predicted branching ratios of $\phi(4S,3D)\to\Lambda\bar\Lambda$ with the new subset of hadronic loops are as follows:
\begin{equation}
\begin{split}
\mathcal{BR}(\phi(4S)\to\Lambda\bar\Lambda)=(1.69-4.99)\times10^{-2}, \\
\mathcal{BR}(\phi(3D)\to\Lambda\bar\Lambda)=(3.33-9.27)\times10^{-2}. \\
\label{branchratio1}
\end{split}
\end{equation}

Strictly speaking, for completeness we should consider the contributions of all possible immediate hadronic loops. However, this is not practicable at this stage, as we are limited by our poor information on the coupling constants of these involved strong interaction vertices. 
In the present subsection, by 
taking $\psi(4S,3D)\to \Lambda\bar{\Lambda}$ as an example, we consider a different subset of the hadronic loops. Our result shown in Eq. (\ref{branchratio1}) is comparable to that in Eq. (\ref{branchratio}). With this effort, we want to show that the treatment presented in Sec. \ref{sec2}, by only considering the subset of hadronic loops composed of $1S$ states of the kaon, can give some indication of the magnitude of the branching ratios discussed.

\section{Predicting the branching ratios of higher strangeonia decays into  $\Sigma\bar\Sigma$ and $\Xi\bar\Xi$}\label{sec3}

\begin{figure}[htbp]\centering
  \includegraphics[width=86mm]{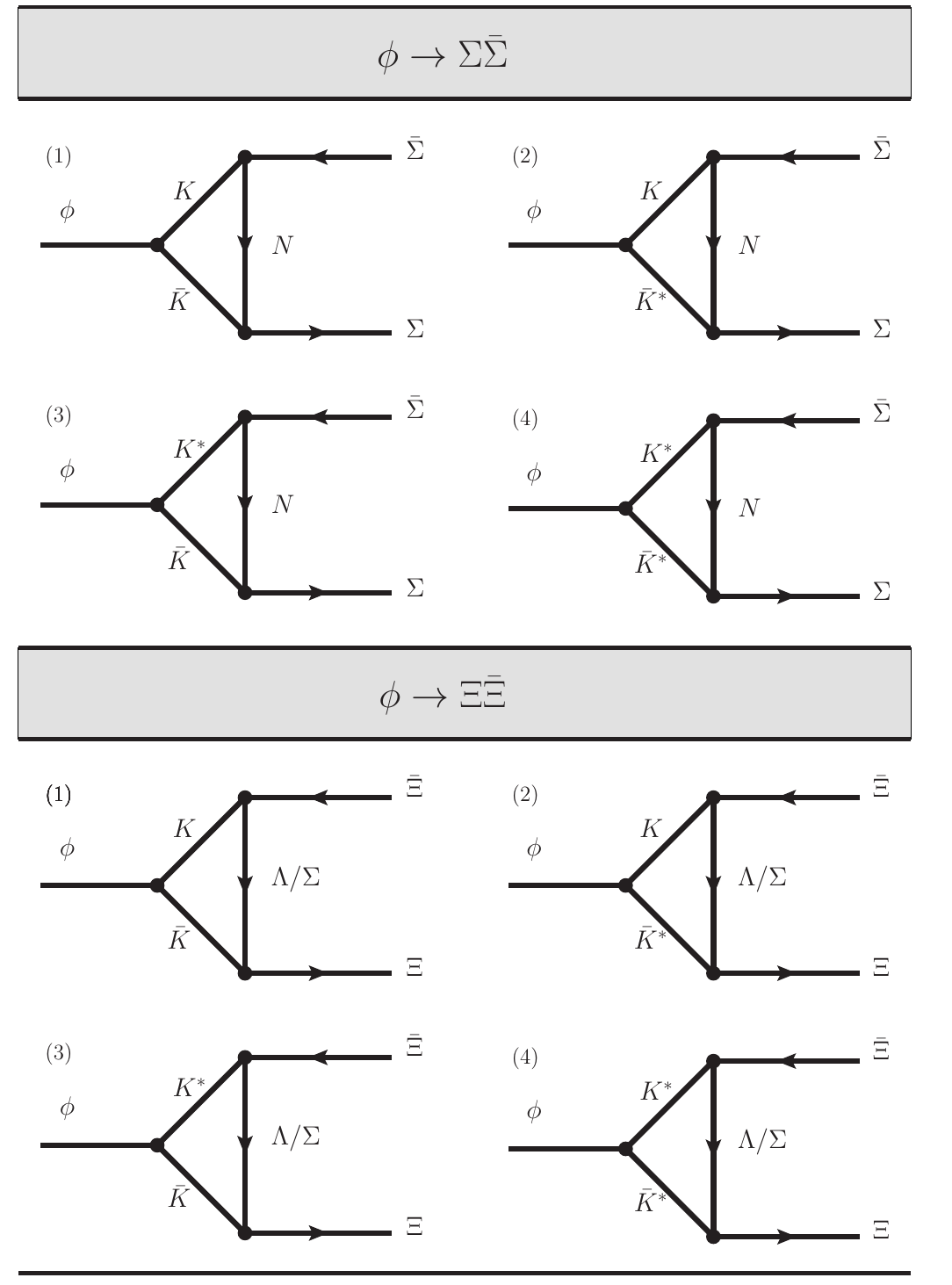}\\
  \caption{The schematic diagrams of the higher strangeonium decays into $\Sigma\bar\Sigma$ and $\Xi\bar{\Xi}$ via the hadronic loop mechanism.}
\label{fig:all}
\end{figure}

We now examine the decays of higher strangeonia into $\Sigma\bar\Sigma$ and $\Xi\bar\Xi$ pairs. Regarding the $\phi\to\Sigma\bar\Sigma$ decays, the schematic diagrams are analogous to those for $\phi\to\Lambda\bar\Lambda$, as depicted in Fig. \ref{fig:all}. The explicit expressions for the decay amplitudes can be found in the Appendix.
In the calculation of the branching ratios for these discussed decays, these coupling constants of the $\Sigma NK^{(*)}$ interactions, 
$f_{\Sigma NK}=0.257$, $g_{\Sigma NK^*}=-2.054$, and $f_{\Sigma NK^*}=0.924
$, are adopted \cite{Nagels:2015lfa}. 

The predicted branching ratios for strangeonia decays into $\Sigma\bar\Sigma$ above the threshold, with $\alpha$ parameter ranging from 3.0 to 4.0, are as follows:
 \begin{equation}
\begin{split}
\mathcal{BR}(\phi(4S)\to\Sigma\bar\Sigma)=(2.44-8.96)\times10^{-5}, \\
\mathcal{BR}(\phi(3D)\to\Sigma\bar\Sigma)=(8.10-30.5)\times10^{-5}, \\
\mathcal{BR}(\phi(5S)\to\Sigma\bar\Sigma)=(3.16-12.0)\times10^{-5}, \\
\mathcal{BR}(\phi(4D)\to\Sigma\bar\Sigma)=(7.46-28.3)\times10^{-5}, \\
\mathcal{BR}(\phi(6S)\to\Sigma\bar\Sigma)=(2.38-9.05)\times10^{-5}, \\
\mathcal{BR}(\phi(5D)\to\Sigma\bar\Sigma)=(5.51-20.9)\times10^{-5}. \\
\end{split}
\end{equation}

In the $\phi\to\Xi\bar\Xi$ decay process, the higher strangeonium $\phi$ initially undergoes a decay into a $K^{(*)}\bar{K}^*$ pair. Subsequently, the $K^{(*)}\bar{K}^*$ pair can transform into a $\Xi\bar\Xi$ pair by exchanging a $\Lambda$ baryon or a $\Sigma$ baryon. The coupling constants used to describe the $\Lambda\Xi K$ and $\Sigma\Xi K$ interactions are \cite{Nagels:2015lfa}
\begin{equation}
\begin{split}
&f_{\Lambda\Xi K}=0.253,\,g_{\Lambda\Xi K^*}=3.557,\,f_{\Lambda\Xi K^*}=6.668,\\
&f_{\Sigma\Xi K}=-0.951,\,g_{\Sigma\Xi K^*}=-2.054,\,f_{\Sigma\Xi K^*}=-13.397.
\end{split}
\end{equation}


By considering the total decay amplitude listed in the Appendix, we can obtain the predicted branching ratios for several higher strangeonia decays into $\Xi\bar\Xi$, with $\alpha$ parameter ranging from 3.0 to 4.0. The results are  \begin{equation}
\begin{split}
\mathcal{BR}(\phi(5S)\to\Xi\bar\Xi)=(3.95-15.4)\times10^{-3}, \\
\mathcal{BR}(\phi(4D)\to\Xi\bar\Xi)=(7.19-29.5)\times10^{-3}, \\
\mathcal{BR}(\phi(6S)\to\Xi\bar\Xi)=(1.18-4.62)\times10^{-2}, \\
\mathcal{BR}(\phi(5D)\to\Xi\bar\Xi)=(1.25-5.20)\times10^{-2}. \\
\end{split}
\end{equation}
These predicted branching ratios can be tested by future experiments like BESIII.

\section{Summary}
\label{sec4}

\textcolor{blue}{}


In this work, we investigate the decays of higher strangeonia into light flavor baryon pairs. By employing the hadronic loop mechanism, we are able to quantitatively calculate their branching ratios, taking into account the unquenched effect which is particularly relevant for higher strangeonia.

Our initial focus is on the decays of higher strangeonia into $\Lambda\bar{\Lambda}$ pairs. The calculation of the branching ratios for these decays is dependent on a parameter $\alpha$ in the dipole form factor. Recently, the BESIII Collaboration provided data on the cross section of $e^+e^-\to \Lambda\bar{\Lambda}$ \cite{BESIII:2023ioy}, allowing us to constrain the range of $\alpha$ by fitting this data. As demonstrated earlier, our results for higher strangeonia decays into $\Lambda\bar{\Lambda}$ agree well with the experimental data, serving as a robust test for the decay mechanism involved in these processes.  

Building on this success, we extend our analysis to predict the decay behaviors of other higher strangeonia decays into baryon pairs such as $\Sigma\bar{\Sigma}$ and $\Xi\bar{\Xi}$, both of which are expected to have sizable branching ratios.

The current experimental capabilities of BESIII \cite{BESIII:2023ioy,BESIII:2017hyw,BESIII:2019nep,BESIII:2020uqk,BESIII:2021rkn,BESIII:2020ktn,BESIII:2021aer,BESIII:2023rwv,BESIII:2022kzc} present a potential opportunity to detect the predicted decay behaviors of the higher strangeonia discussed in this study. The validation of our predictions by experimental observations will be an interesting task in the near future, providing valuable insights into the decays of these higher strangeonia.

\begin{acknowledgements}
This work is supported by the China National Funds for Distinguished Young Scientists under  Grant No. 11825503, the National Key Research and Development Program of China under Contract No. 2020YFA0406400, the 111 Project under Grant
No. B20063, the National Natural Science Foundation of China under Grants No. 12247101 and No. 12335001,
the Project for Top-notch Innovative Talents of Gansu Province, and the fundamental Research Funds for the Central Universities. Z.Y. Bai is also supported by Education Department of Gansu Province for Excellent Graduate Student ``Innovation Star" under Grant No. 2022CXZX-060.
\end{acknowledgements}

\appendix
\section*{APPENDIX: Decay amplitudes for higher strangeonium decays into $\Sigma\bar\Sigma$ and $\Xi\bar{\Xi}$}
\label{app01}
\setcounter{equation}{0}
\renewcommand\theequation{A\arabic{equation}} 
We present the concrete expressions of the decay amplitudes of $\phi\to\Sigma\bar\Sigma$ and $\phi\to\Xi\bar\Xi$ transitions in this Appendix.

The decay amplitudes of $\phi\to\Sigma\bar\Sigma$ as shown in Fig. \ref{fig:all} are
\begin{equation}
\begin{split}
\mathcal{M}_{(1)}^{\Sigma\bar\Sigma}=&i^3\int\frac{dq^4}{(2\pi)^4}g_{\phi KK}\epsilon_\phi^\mu(q_{1\mu}- q_{2\mu})\left(\frac{f_{\Sigma NK}}{m_{\pi^+}}\right)^2  \\
&\times\bar u(p_2)\slashed{q}_2\gamma^5(\slashed q+m_N)\slashed q_1\gamma^5v(p_1)\\
&\times\frac{1}{q_1^2-m_K^2}\frac{1}{q_2^2-m_K^2}\frac{1}{q^2-m_N^2}\mathcal{F}^{2}(q^2,m_N^2),
\end{split}
\end{equation}

\begin{equation}
\begin{split}
\mathcal{M}_{(2)}^{\Sigma\bar\Sigma}=&-i^3\int\frac{dq^4}{(2\pi)^4}g_{\phi KK^*}\varepsilon_{\mu\nu\alpha\beta}\epsilon_\phi^\mu p^\nu q_2^\alpha\bar u(p_2)\Big(g_{\Sigma NK^*}\gamma_\rho \\
&+\frac{f_{\Sigma NK^*}}{2m_p}q_2^\lambda\sigma_{\lambda\rho}\Big)(\slashed q+m_N)\frac{f_{\Sigma NK}}{m_{\pi^+}}\slashed q_1\gamma^5v(p_1)\\
&\times\frac{1}{q_1^2-m_K^2}\frac{-g^{\beta\rho}+q_2^\beta q_2^\rho\slash{m_{K^{*}}^2}}{q_2^2-m_{K^{*}}^{2}}\frac{1}{q^2-m_N^2}\mathcal{F}^{2}(q^2,m_N^2),\\
\end{split}
\end{equation}

\begin{equation}
\begin{split}
\mathcal{M}_{(3)}^{\Sigma\bar\Sigma}=&-i^3\int\frac{dq^4}{(2\pi)^4}g_{\phi KK^*}\varepsilon_{\mu\nu\alpha\beta}\epsilon_\phi^\mu p^\nu q_1^\alpha\bar u(p_2)\frac{f_{\Sigma NK}}{m_{\pi^+}}\slashed q_2\gamma^5\\
&\times(\slashed q+m_N)\Big(g_{\Sigma NK^*}\gamma_\rho+\frac{f_{\Sigma NK^*}}{2m_p}q_1^\lambda\sigma_{\lambda\rho}\Big)v(p_1)\\
&\times\frac{-g^{\beta\rho}+q_1^\beta q_1^\rho\slash{m_{K^{*}}^2}}{q_1^2-m_{K^{*}}^{2}}\frac{1}{q_2^2-m_K^{2}}\frac{1}{q^2-m_N^2}\mathcal{F}^{2}(q^2,m_N^2),\\
\end{split}
\end{equation}

\begin{equation}
\begin{split}
\mathcal{M}_{(4)}^{\Sigma\bar\Sigma}=&i^3\int\frac{dq^4}{(2\pi)^4}g_{\phi K^*K^*}\epsilon_\phi^\mu\Big(g_{\alpha\beta}(q_{1\mu}-q_{2\mu})+g_{\beta\mu}(p_\alpha+q_{2\alpha})\\
&-g_{\alpha\mu}(p_\beta+q_{1\beta})\Big)\bar u(p_2)\Big(g_{\Sigma NK^*}\gamma_\rho+\frac{f_{\Sigma NK^*}}{2m_p}q_2^\lambda\sigma_{\lambda\rho}\Big)\\
&\times(\slashed q+m_N)\Big(g_{\Sigma NK^*}\gamma_\eta+\frac{f_{\Sigma NK^*}}{2m_p}q_1^\kappa\sigma_{\kappa\eta}\Big)v(p_1) \\
&\times\frac{-g^{\alpha\eta}+q_1^\alpha q_1^\eta\slash{m_{K^{*}}^2}}{q_1^2-m_{K^{*}}^{2}}\frac{-g^{\beta\rho}+q_2^\beta q_2^\rho\slash{m_{K^{*}}^2}}{q_2^2-m_{K^{*}}^{2}}\frac{1}{q^2-m_N^2}\\
&\times\mathcal{F}^{2}(q^2,m_N^2).\\
\end{split}
\end{equation}

The total decay amplitude of $\phi\to\Sigma\bar\Sigma$ can be expressed as
\begin{equation}
\mathcal{M}_{\text{total}}^{\Sigma\bar\Sigma}=2(\mathcal{M}_{(1)}^{\Sigma\bar\Sigma}+\mathcal{M}_{(2)}^{\Sigma\bar\Sigma}+\mathcal{M}_{(3)}^{\Sigma\bar\Sigma}+\mathcal{M}_{(4)}^{\Sigma\bar\Sigma}).
\end{equation}

The decay amplitudes of $\phi\to\Xi\bar\Xi$ via exchanging the $\Lambda$ baryon as shown in Fig. \ref{fig:all} are
\begin{equation}
\begin{split}
\mathcal{M}_{\Lambda(1)}^{\Xi\bar\Xi}=&i^3\int\frac{dq^4}{(2\pi)^4}g_{\phi KK}\epsilon_\phi^\mu(q_{1\mu}-q_{2\mu})\left(\frac{f_{\Lambda\Xi K}}{m_{\pi^+}}\right)^2  \\
&\times\bar u(p_2)\slashed{q}_2\gamma^5(\slashed q+m_\Lambda)\slashed q_1\gamma^5v(p_1)\\
&\times\frac{1}{q_1^2-m_K^2}\frac{1}{q_2^2-m_K^2}\frac{1}{q^2-m_\Lambda^2}\mathcal{F}^{2}(q^2,m_\Lambda^2),
\end{split}
\end{equation}

\begin{equation}
\begin{split}
\mathcal{M}_{\Lambda(2)}^{\Xi\bar\Xi}=&-i^3\int\frac{dq^4}{(2\pi)^4}g_{\phi KK^*}\varepsilon_{\mu\nu\alpha\beta}\epsilon_\phi^\mu p^\nu q_2^\alpha\bar u(p_2)\Big(g_{\Lambda\Xi K^*}\gamma_\rho \\
&+\frac{f_{\Lambda\Xi K^*}}{2m_p}q_2^\lambda\sigma_{\lambda\rho}\Big)(\slashed q+m_\Lambda)\frac{f_{\Lambda\Xi K}}{m_{\pi^+}}\slashed q_1\gamma^5v(p_1)\\
&\times\frac{1}{q_1^2-m_K^2}\frac{-g^{\beta\rho}+q_2^\beta q_2^\rho\slash{m_{K^{*}}^2}}{q_2^2-m_{K^{*}}^{2}}\frac{1}{q^2-m_\Lambda^2}\mathcal{F}^{2}(q^2,m_\Lambda^2),\\
\end{split}
\end{equation}

\begin{equation}
\begin{split}
\mathcal{M}_{\Lambda(3)}^{\Xi\bar\Xi}=&-i^3\int\frac{dq^4}{(2\pi)^4}g_{\phi KK^*}\varepsilon_{\mu\nu\alpha\beta}\epsilon_\phi^\mu p^\nu q_1^\alpha\bar u(p_2)\frac{f_{\Lambda\Xi K}}{m_{\pi^+}}\slashed q_2\gamma^5\\
&\times(\slashed q+m_\Lambda)\Big(g_{\Lambda\Xi K^*}\gamma_\rho+\frac{f_{\Lambda\Xi K^*}}{2m_p}q_1^\lambda\sigma_{\lambda\rho}\Big)v(p_1)\\
&\times\frac{-g^{\beta\rho}+q_1^\beta q_1^\rho\slash{m_{K^{*}}^2}}{q_1^2-m_{K^{*}}^{2}}\frac{1}{q_2^2-m_K^{2}}\frac{1}{q^2-m_\Lambda^2}\mathcal{F}^{2}(q^2,m_\Lambda^2),\\
\end{split}
\end{equation}

\begin{equation}
\begin{split}
\mathcal{M}_{\Lambda(4)}^{\Xi\bar\Xi}=&i^3\int\frac{dq^4}{(2\pi)^4}g_{\phi K^*K^*}\epsilon_\phi^\mu\Big(g_{\alpha\beta}(q_{1\mu}-q_{2\mu})+g_{\beta\mu}(p_\alpha+q_{2\alpha})\\
&-g_{\alpha\mu}(p_\beta+q_{1\beta})\Big)\bar u(p_2)\Big(g_{\Lambda\Xi K^*}\gamma_\rho+\frac{f_{\Lambda\Xi K^*}}{2m_p}q_2^\lambda\sigma_{\lambda\rho}\Big)\\
&\times(\slashed q+m_\Lambda)\Big(g_{\Lambda\Xi K^*}\gamma_\eta+\frac{f_{\Lambda\Xi K^*}}{2m_p}q_1^\kappa\sigma_{\kappa\eta}\Big)v(p_1) \\
&\times\frac{-g^{\alpha\eta}+q_1^\alpha q_1^\eta\slash{m_{K^{*}}^2}}{q_1^2-m_{K^{*}}^{2}}\frac{-g^{\beta\rho}+q_2^\beta q_2^\rho\slash{m_{K^{*}}^2}}{q_2^2-m_{K^{*}}^{2}}\frac{1}{q^2-m_\Lambda^2}\\
&\times\mathcal{F}^{2}(q^2,m_\Lambda^2).\\
\end{split}
\end{equation}

The decay amplitudes of $\phi\to\Xi\bar\Xi$ via exchanging the $\Sigma$ baryon as shown in Fig. \ref{fig:all} are
\begin{equation}
\begin{split}
\mathcal{M}_{\Sigma(1)}^{\Xi\bar\Xi}=&i^3\int\frac{dq^4}{(2\pi)^4}g_{\phi KK}\epsilon_\phi^\mu(q_{1\mu}-q_{2\mu})\left(\frac{f_{\Sigma\Xi K}}{m_{\pi^+}}\right)^2  \\
&\times\bar u(p_2)\slashed{q}_2\gamma^5(\slashed q+m_\Sigma)\slashed q_1\gamma^5v(p_1)\\
&\times\frac{1}{q_1^2-m_K^2}\frac{1}{q_2^2-m_K^2}\frac{1}{q^2-m_\Sigma^2}\mathcal{F}^{2}(q^2,m_\Sigma^2),
\end{split}
\end{equation}

\begin{equation}
\begin{split}
\mathcal{M}_{\Sigma(2)}^{\Xi\bar\Xi}=&-i^3\int\frac{dq^4}{(2\pi)^4}g_{\phi KK^*}\varepsilon_{\mu\nu\alpha\beta}\epsilon_\phi^\mu p^\nu q_2^\alpha\bar u(p_2)\Big(g_{\Sigma\Xi K^*}\gamma_\rho \\
&+\frac{f_{\Sigma\Xi K^*}}{2m_p}q_2^\lambda\sigma_{\lambda\rho}\Big)(\slashed q+m_\Sigma)\frac{f_{\Sigma\Xi K}}{m_{\pi^+}}\slashed q_1\gamma^5v(p_1)\\
&\times\frac{1}{q_1^2-m_K^2}\frac{-g^{\beta\rho}+q_2^\beta q_2^\rho\slash{m_{K^{*}}^2}}{q_2^2-m_{K^{*}}^{2}}\frac{1}{q^2-m_\Sigma^2}\mathcal{F}^{2}(q^2,m_\Sigma^2),\\
\end{split}
\end{equation}

\begin{equation}
\begin{split}
\mathcal{M}_{\Sigma(3)}^{\Xi\bar\Xi}=&-i^3\int\frac{dq^4}{(2\pi)^4}g_{\phi KK^*}\varepsilon_{\mu\nu\alpha\beta}\epsilon_\phi^\mu p^\nu q_1^\alpha\bar u(p_2)\frac{f_{\Sigma\Xi K}}{m_{\pi^+}}\slashed q_2\gamma^5\\
&\times(\slashed q+m_\Sigma)\Big(g_{\Sigma\Xi K^*}\gamma_\rho+\frac{f_{\Sigma\Xi K^*}}{2m_p}q_1^\lambda\sigma_{\lambda\rho}\Big)v(p_1)\\
&\times\frac{-g^{\beta\rho}+q_1^\beta q_1^\rho\slash{m_{K^{*}}^2}}{q_1^2-m_{K^{*}}^2}\frac{1}{q_2^2-m_K^{2}}\frac{1}{q^2-m_\Sigma^2}\mathcal{F}^{2}(q^2,m_\Sigma^2),\\
\end{split}
\end{equation}

\begin{equation}
\begin{split}
\mathcal{M}_{\Sigma(4)}^{\Xi\bar\Xi}=&i^3\int\frac{dq^4}{(2\pi)^4}g_{\phi K^*K^*}\epsilon_\phi^\mu\Big(g_{\alpha\beta}(q_{1\mu}-q_{2\mu})+g_{\beta\mu}(p_\alpha+q_{2\alpha})\\
&-g_{\alpha\mu}(p_\beta+q_{1\beta})\Big)\bar u(p_2)\Big(g_{\Sigma\Xi K^*}\gamma_\rho+\frac{f_{\Sigma\Xi K^*}}{2m_p}q_2^\lambda\sigma_{\lambda\rho}\Big)\\
&\times(\slashed q+m_\Sigma)\Big(g_{\Sigma\Xi K^*}\gamma_\eta+\frac{f_{\Sigma\Xi K^*}}{2m_p}q_1^\kappa\sigma_{\kappa\eta}\Big)v(p_1) \\
&\times\frac{-g^{\alpha\eta}+q_1^\alpha q_1^\eta\slash{m_{K^{*}}^2}}{q_1^2-m_{K^{*}}^{2}}\frac{-g^{\beta\rho}+q_2^\beta q_2^\rho\slash{m_{K^{*}}^2}}{q_2^2-m_{K^{*}}^{2}}\frac{1}{q^2-m_\Sigma^2}\\
&\times\mathcal{F}^{2}(q^2,m_\Sigma^2).
\end{split}
\end{equation}

The total decay amplitude of $\phi\to\Xi\bar\Xi$ can be expressed as
\begin{equation}
\mathcal{M}_{\text{total}}^{\Xi\bar{\Xi}}=2\left(\mathcal{M}_\Lambda^{\Xi\bar\Xi}+\mathcal{M}_\Sigma^{\Xi\bar\Xi}\right)
\end{equation}
with
\begin{equation}
\begin{split}
\mathcal{M}_\Lambda^{\Xi\bar\Xi}&=\mathcal{M}_{\Lambda(1)}^{\Xi\bar\Xi}+\mathcal{M}_{\Lambda(2)}^{\Xi\bar\Xi}+\mathcal{M}_{\Lambda(3)}^{\Xi\bar\Xi}+\mathcal{M}_{\Lambda(4)}^{\Xi\bar\Xi},\\
\mathcal{M}_\Sigma^{\Xi\bar\Xi}&=\mathcal{M}_{\Sigma(1)}^{\Xi\bar\Xi}+\mathcal{M}_{\Sigma(2)}^{\Xi\bar\Xi}+\mathcal{M}_{\Sigma(3)}^{\Xi\bar\Xi}+\mathcal{M}_{\Sigma(4)}^{\Xi\bar\Xi}.
\end{split}
\end{equation}

The decay amplitudes of $\phi\to\Lambda\bar\Lambda$ corresponding to Fig. \ref{fig:K1}
are shown as

\begin{equation}
\begin{split}
\mathcal{M}_{(5)}^{\Lambda\bar{\Lambda}}=&i^3\int\frac{dq^4}{(2\pi)^4}g_{\phi KK_1(1270)}\epsilon_\phi^\mu\bar u(p_2)  \\
&\times\Big(-\frac{if_{N\Lambda K_1}}{m_{b_1}}\cos\theta\, q_2^\lambda\sigma_{\rho\lambda}\gamma^5+\sin\theta\,(g_{N\Lambda K_1^\prime}\gamma_\rho\gamma^5\\
&-\frac{if_{N\Lambda K_1^\prime}}{m_p}q_{2\rho}\gamma^5)\Big)(\slashed q+m_N)\Big(\frac{f_{N\Lambda K}}{m_{\pi^+}}\slashed q_1\gamma^5\Big)v(p_1)\\
&\times\frac{1}{q_1^2-m_K^2}\frac{-g^{\mu\rho}+q_2^\mu q_2^\rho/m_{K_1(1270)}^2}{q_2^2-m_{K_1(1270)}^2}\frac{1}{q^2-m_N^2}\mathcal{F}^{2}(q^2,m_N^2),
\end{split}
\end{equation}

\begin{equation}
\begin{split}
\mathcal{M}_{(6)}^{\Lambda\bar{\Lambda}}=&i^3\int\frac{dq^4}{(2\pi)^4}g_{\phi KK_1(1270)}\epsilon_\phi^\mu\bar u(p_2)  \\
&\times\Big(\frac{f_{N\Lambda K}}{m_{\pi^+}}\slashed q_2\gamma^5\Big)(\slashed q+m_N)\Big(-\frac{if_{N\Lambda K_1}}{m_{b_1}}\cos\theta\, q_1^\lambda\sigma_{\rho\lambda}\gamma^5\\
&+\sin\theta\,(g_{N\Lambda K_1^\prime}\gamma_\rho\gamma^5-\frac{if_{N\Lambda K_1^\prime}}{m_p}q_{1\rho}\gamma^5)\Big)v(p_1)\\
&\times\frac{-g^{\mu\rho}+q_1^\mu q_1^\rho/m_{K_1(1270)}^2}{q_1^2-m_{K_1(1270)}^2}\frac{1}{q_2^2-m_K^2}\frac{1}{q^2-m_N^2}\mathcal{F}^{2}(q^2,m_N^2),
\end{split}
\end{equation}

\begin{equation}
\begin{split}
\mathcal{M}_{(7)}^{\Lambda\bar{\Lambda}}=&i^3\int\frac{dq^4}{(2\pi)^4}g_{\phi K^*K_1(1270)}\varepsilon_{\mu\nu\alpha\beta}\epsilon_\phi^\mu( p^2q_1^\nu+q_1^2p^\nu)\bar u(p_2)\\
&\times\Big(-\frac{if_{N\Lambda K_1}}{m_{b_1}}\cos\theta\, q_2^\lambda\sigma_{\rho\lambda}\gamma^5+\sin\theta \,(g_{N\Lambda K_1^\prime}\gamma_\rho\gamma^5\\
&-\frac{if_{N\Lambda K_1^\prime}}{m_p}q_{2\rho}\gamma^5)\Big)(\slashed q+m_N)\Big(g_{\Lambda NK^*}\gamma_\omega\\
&+\frac{f_{\Lambda NK^*}}{2m_p}q_1^\phi\sigma_{\phi\omega}\Big)v(p_1)\frac{g^{\alpha\omega}+q_1^\alpha q_1^\omega/m_{K^*}^2}{q_1^2-m_{K^*}^2}\\
&\times\frac{g^{\beta\rho}+q_2^\beta q_2^\rho/m_{K_1(1270)}^2}{q_2^2-m_{K_1(1270)}^2}\frac{1}{q^2-m_N^2}\mathcal{F}^{2}(q^2,m_N^2),
\end{split}
\end{equation}

\begin{equation}
\begin{split}
\mathcal{M}_{(8)}^{\Lambda\bar{\Lambda}}=&-i^3\int\frac{dq^4}{(2\pi)^4}g_{\phi K^*K_1(1270)}\varepsilon_{\mu\nu\alpha\beta}\epsilon_\phi^\mu( p^2q_2^\nu+q_2^2p^\nu)\bar u(p_2)\\
&\times\Big(g_{\Lambda NK^*}\gamma_\omega+\frac{f_{\Lambda NK^*}}{2m_p}q_1^\phi\sigma_{\phi\omega}\Big)(\slashed q+m_N)\\
&\times\Big(-\frac{if_{N\Lambda K_1}}{m_{b_1}}\cos\theta\, q_1^\lambda\sigma_{\rho\lambda}\gamma^5+\sin\theta\,(g_{N\Lambda K_1^\prime}\gamma_\rho\gamma^5\\
&-\frac{if_{N\Lambda K_1^\prime}}{m_p}q_{1\rho}\gamma^5)\Big)\frac{g^{\beta\rho}+q_1^\beta q_1^\rho/m_{K_1(1270)}^2}{q_1^2-m_{K_1(1270)}^2}\\
&\times\frac{g^{\alpha\omega}+q_2^\alpha q_2^\omega/m_{K^*}^2}{q_2^2-m_{K^*}^2}\frac{1}{q^2-m_N^2}\mathcal{F}^{2}(q^2,m_N^2),
\end{split}
\end{equation}

\begin{equation}
\begin{split}
\mathcal{M}_{(9)}^{\Lambda\bar{\Lambda}}=&i^3\int\frac{dq^4}{(2\pi)^4}g_{\phi KK_1(1400)}\epsilon_\phi^\mu\bar u(p_2)  \\
&\times\Big(\frac{if_{N\Lambda K_1}}{m_{b_1}}\sin\theta\, q_2^\lambda\sigma_{\rho\lambda}\gamma^5+\cos\theta\,(g_{N\Lambda K_1^\prime}\gamma_\rho\gamma^5\\
&-\frac{if_{N\Lambda K_1^\prime}}{m_p}q_{2\rho}\gamma^5)\Big)(\slashed q+m_N)\Big(\frac{f_{N\Lambda K}}{m_{\pi^+}}\slashed q_1\gamma^5\Big)v(p_1)\\
&\times\frac{1}{q_1^2-m_K^2}\frac{-g^{\mu\rho}+q_2^\mu q_2^\rho/m_{K_1(1400)}^2}{q_2^2-m_{K_1(1400)}^2}\frac{1}{q^2-m_N^2}\mathcal{F}^{2}(q^2,m_N^2),
\end{split}
\end{equation}

\begin{equation}
\begin{split}
\mathcal{M}_{(10)}^{\Lambda\bar{\Lambda}}=&i^3\int\frac{dq^4}{(2\pi)^4}g_{\phi KK_1(1400)}\epsilon_\phi^\mu\bar u(p_2)  \\
&\times\Big(\frac{f_{N\Lambda K}}{m_{\pi^+}}\slashed q_2\gamma^5\Big)(\slashed q+m_N)\Big(\frac{if_{N\Lambda K_1}}{m_{b_1}}\sin\theta\, q_1^\lambda\sigma_{\rho\lambda}\gamma^5\\
&+\cos\theta\,(g_{N\Lambda K_1^\prime}\gamma_\rho\gamma^5-\frac{if_{N\Lambda K_1^\prime}}{m_p}q_{1\rho}\gamma^5)\Big)v(p_1)\\
&\times\frac{-g^{\mu\rho}+q_1^\mu q_1^\rho/m_{K_1(1400)}^2}{q_1^2-m_{K_1(1400)}^2}\frac{1}{q_2^2-m_K^2}\frac{1}{q^2-m_N^2}\mathcal{F}^{2}(q^2,m_N^2),
\end{split}
\end{equation}

\begin{equation}
\begin{split}
\mathcal{M}_{(11)}^{\Lambda\bar{\Lambda}}=&i^3\int\frac{dq^4}{(2\pi)^4}g_{\phi K^*K_1(1400)}\varepsilon_{\mu\nu\alpha\beta}\epsilon_\phi^\mu( p^2q_1^\nu+q_1^2p^\nu)\bar u(p_2)\\
&\times\Big(\frac{if_{N\Lambda K_1}}{m_{b_1}}\sin\theta\, q_2^\lambda\sigma_{\rho\lambda}\gamma^5+\cos\theta\,(g_{N\Lambda K_1^\prime}\gamma_\rho\gamma^5\\
&-\frac{if_{N\Lambda K_1^\prime}}{m_p}q_{2\rho}\gamma^5)\Big)(\slashed q+m_N)\Big(g_{\Lambda NK^*}\gamma_\omega\\
&+\frac{f_{\Lambda NK^*}}{2m_p}q_1^\phi\sigma_{\phi\omega}\Big)v(p_1)\frac{g^{\alpha\omega}+q_1^\alpha q_1^\omega/m_{K^*}^2}{q_1^2-m_{K^*}^2}\\
&\times\frac{g^{\beta\rho}+q_2^\beta q_2^\rho/m_{K_1(1400)}^2}{q_2^2-m_{K_1(1400)}^2}\frac{1}{q^2-m_N^2}\mathcal{F}^{2}(q^2,m_N^2),
\end{split}
\end{equation}

\begin{equation}
\begin{split}
\mathcal{M}_{(12)}^{\Lambda\bar{\Lambda}}=&-i^3\int\frac{dq^4}{(2\pi)^4}g_{\phi K^*K_1(1400)}\varepsilon_{\mu\nu\alpha\beta}\epsilon_\phi^\mu( p^2q_2^\nu+q_2^2p^\nu)\bar u(p_2)\\
&\times\Big(g_{\Lambda NK^*}\gamma_\omega+\frac{f_{\Lambda NK^*}}{2m_p}q_1^\phi\sigma_{\phi\omega}\Big)(\slashed q+m_N)\\
&\times\Big(\frac{if_{N\Lambda K_1}}{m_{b_1}}\sin\theta\, q_1^\lambda\sigma_{\rho\lambda}\gamma^5+\cos\theta\,(g_{N\Lambda K_1^\prime}\gamma_\rho\gamma^5\\
&-\frac{if_{N\Lambda K_1^\prime}}{m_p}q_{1\rho}\gamma^5)\Big)\frac{g^{\beta\rho}+q_1^\beta q_1^\rho/m_{K_1(1400)}^2}{q_1^2-m_{K_1(1400)}^2}\\
&\times\frac{g^{\alpha\omega}+q_2^\alpha q_2^\omega/m_{K^*}^2}{q_2^2-m_{K^*}^2}\frac{1}{q^2-m_N^2}\mathcal{F}^{2}(q^2,m_N^2).
\end{split}
\end{equation}

The total amplitude of $\phi\to\Lambda\bar\Lambda$ via another subset as discussed in Sec. \ref{d} can be expressed as
\begin{equation}
\begin{split}
\mathcal{M}_{\text{total}}^{\Lambda\bar{\Lambda}} =&2\Big(\mathcal{M}_{(1)}^{\Lambda\bar{\Lambda}} + \mathcal{M}_{(2)}^{\Lambda\bar{\Lambda}} + \mathcal{M}_{(3)}^{\Lambda\bar{\Lambda}} + \mathcal{M}_{(4)}^{\Lambda\bar{\Lambda}}\\
&+\mathcal{M}_{(5)}^{\Lambda\bar{\Lambda}} + \mathcal{M}_{(6)}^{\Lambda\bar{\Lambda}} + \mathcal{M}_{(7)}^{\Lambda\bar{\Lambda}} + \mathcal{M}_{(8)}^{\Lambda\bar{\Lambda}}\\
&+\mathcal{M}_{(9)}^{\Lambda\bar{\Lambda}} + \mathcal{M}_{(10)}^{\Lambda\bar{\Lambda}} + \mathcal{M}_{(11)}^{\Lambda\bar{\Lambda}} + \mathcal{M}_{(12)}^{\Lambda\bar{\Lambda}}\Big).
\end{split}
\end{equation}
~\\
~\\
~\\
~\\
~\\
~\\
~\\
~\\
~\\
~\\
~\\
~\\
~\\
~\\
~\\
~\\
~\\


\end{document}